\documentclass[11pt]{article}

\usepackage[margin=1in]{geometry}
\usepackage{amsmath,amssymb,amsfonts}
\usepackage{graphicx}
\usepackage{booktabs}
\usepackage{siunitx}
\usepackage{xcolor}
\usepackage[round]{natbib}
\usepackage{hyperref}
\hypersetup{
    colorlinks=true,
    linkcolor=blue!50!black,
    citecolor=green!40!black,
    urlcolor=blue!60!black
}
\usepackage{authblk}

\newcommand{\Bmag}{B}
\newcommand{\STFT}{\mathcal{S}}
\newcommand{\fs}{f_{s}}

\title{\bfseries Cross-Sensor RGB Spectrograms:\\
       A Visual Method for Anomaly Detection in\\
       Classical and Quantum Magnetometer Triads}

\author[1]{Manas Pandey}
\affil[1]{Indian Institute of Technology, Kanpur, Uttar Pradesh 208016, India}

\date{\today}

\begin{document}
\maketitle

\begin{abstract}
\noindent
Stationary multi-magnetometer arrays are routinely deployed in
geomagnetic observatories, laboratory shielded rooms, and
ground-based monitoring stations.  The standard analysis pipeline
reduces each sensor to an independent power spectrum, discarding any
inter-sensor structure that is itself diagnostic of measurement
health and of localised magnetic activity.  This paper develops a
purely theoretical framework for a deliberately simple visualisation
that maps the short-time Fourier (STFT) power spectra of three
concurrent magnetometers into the red, green, and blue channels of a
single image: the \emph{cross-sensor RGB spectrogram}.  Inter-sensor
coherence appears as neutral grey or white, while spectral energy
that is unique to one or two sensors stands out as saturated colour.
We formalise the construction of the image, derive its
time--frequency resolution properties, give an explicit account of
the per-channel normalisation choice, and present a colour--anomaly
taxonomy that distinguishes coherent broadband activity, single-sensor
faults, asymmetric pairwise sources, and slow temporal drift.  A
companion long-window variant is described for resolving features in
the ultra-low frequency (ULF) band.  The construction is presented
without reference to any particular dataset or implementation; it is
intended as a self-contained methodological building block that can
be inserted into any monitoring pipeline whose front end is a
synchronously sampled magnetometer triad.  Because the construction
operates on scalar magnitude time series alone, it applies equally to
classical fluxgate sensors and to quantum magnetometers---optically
pumped magnetometers (OPMs), nitrogen-vacancy (NV) centre arrays, and
superconducting quantum interference devices (SQUIDs)---where
distinguishing quantum-limited noise from technical artefacts is a
central diagnostic challenge.
\end{abstract}

\medskip
\noindent\textbf{Keywords:} magnetometer triad, cross-sensor analysis,
short-time Fourier transform, RGB visualisation, anomaly detection,
ultra-low frequency, time--frequency analysis, quantum magnetometry,
quantum sensing.

\section{Introduction}
\label{sec:intro}

Stationary magnetometer arrays underlie a broad class of measurement
problems, from monitoring geomagnetic activity at ground observatories,
to maintaining the stability of magnetically shielded laboratories,
to characterising localised disturbances in industrial environments.
A frequent design choice is to deploy three colocated or
near-colocated sensors, both for redundancy against the failure of any
one channel and to enable inter-sensor consistency checks that single
instruments cannot provide on their own.

The emergence of quantum magnetometers---optically pumped
magnetometers (OPMs) based on spin-polarised alkali vapours, diamond
nitrogen-vacancy (NV) centre sensors, and superconducting quantum
interference devices (SQUIDs)---has extended precision magnetometry
into regimes where the noise floor is set by fundamental quantum
processes such as spin-projection noise and photon shot noise
\citep{budker2007optical, degen2017quantum}.  Triad configurations of
such sensors are now common in magnetically shielded rooms and in
searches for exotic spin-dependent interactions.  In these settings
the diagnostic question is sharpened: an anomaly that appears in one
sensor may be a quantum-limited fluctuation, a technical artefact, or
a genuine external signal, and the answer often hinges on whether the
other two sensors see it as well.

The standard frequency-domain analysis of such an array proceeds one
sensor at a time: a power spectral density per sensor, occasionally a
pairwise coherence between two sensors, but rarely a joint
visualisation of all three.  This leaves a substantial blind spot.
Sensor faults that affect a single channel, narrow-band
electromagnetic interference (EMI) from nearby equipment, and
asymmetric magnetic disturbances all imprint themselves
\emph{differentially} across the array, and are most quickly
recognised by an analyst when the comparison is made visually rather
than by tabulation.

This paper introduces a deliberately simple construction that turns
three concurrent scalar magnetometer streams into a single colour
image: the \emph{cross-sensor RGB spectrogram}.  The red, green, and
blue channels of the output image carry, respectively, the
normalised short-time Fourier (STFT) power of the first, second, and
third sensor of the triad.  Wherever the three sensors agree on the
spectral content of the signal, the resulting pixel is achromatic
(grey or white); wherever a single sensor dominates, the pixel is
saturated in the corresponding primary colour; wherever two sensors
agree but the third dissents, a secondary colour (cyan, magenta, or
yellow) appears.  Anomalies thus become both spatially and
chromatically localised in the time--frequency plane in a way that is
immediately legible to a human analyst and that is also amenable to
downstream quantitative post-processing.

The contribution of this work is theoretical and is fourfold:

\begin{enumerate}
    \item A formal definition of the cross-sensor RGB spectrogram,
          including its windowing, normalisation, and long-window
          low-frequency variant;
    \item An analytical account of the resulting time--frequency
          resolution and its trade-offs as a function of the
          analysis-window length and overlap;
    \item A catalogue of colour signatures and their physical
          interpretation, distinguishing sensor-side from
          source-side anomalies; and
    \item A discussion of the visualisation's relationship to
          classical pairwise coherence and of its principled
          limitations.
\end{enumerate}

The paper is organised as follows.  Section~\ref{sec:related}
reviews relevant prior art.  Section~\ref{sec:setup} fixes notation
for a stationary magnetometer triad.  Section~\ref{sec:method}
formalises the construction.  Section~\ref{sec:signatures} gives the
colour--anomaly taxonomy.  Section~\ref{sec:discussion} treats the
relationship to coherence, normalisation choices, window-length
sensitivity, and limitations, and Section~\ref{sec:conclusion}
concludes.

\section{Related Work}
\label{sec:related}

\paragraph{Time--frequency analysis.}
The short-time Fourier transform (STFT) and its windowed descendants
form the canonical tool for non-stationary signal analysis
\citep{allen1977unified, cohen1995time, oppenheim1999discrete}.  The
choice of analysis window has been the subject of a vast literature;
\citet{harris1978windows} remains the standard reference for the
trade-offs among rectangular, Hann, Hamming, Blackman--Harris, and
related families.  Welch's method \citep{welch1967psd} provides the
dominant variance-reduction approach when only stationary spectra
are required.

\paragraph{False-colour visualisation.}
The use of three independent scalar fields as the R, G, and B
channels of a colour image has a long history in remote sensing,
where multispectral data are routinely composited into ``false-colour''
images to expose differences invisible to monochrome inspection
\citep{gonzalez2018digital}.  The construction proposed here imports
the same pictorial idea into the time--frequency analysis of a
three-element magnetometer array.

\paragraph{Geomagnetic micropulsations.}
Variations of the geomagnetic field at very low frequencies have
been classified into Pc1--Pc5 (continuous) and Pi1--Pi2 (irregular)
bands since \citet{jacobs1964classification}.  These bands lie
almost entirely below 1\,\si{Hz}, which motivates the long-window
low-frequency variant of the construction described in
Section~\ref{sec:lowf}.

\paragraph{Quantum magnetometry.}
Comprehensive reviews of quantum sensing are given by
\citet{budker2007optical} for optical magnetometry and by
\citet{degen2017quantum} for the broader quantum-sensing framework.
The noise processes that limit quantum magnetometers---spin-projection
noise, photon shot noise, SQUID flux noise, and NV-centre
$T_{1}$/$T_{2}$ relaxation---have well-characterised spectral
signatures that are distinct from classical technical noise.  We do
not develop a quantum noise model here, but we note in
Section~\ref{sec:quantumsig} how these signatures map onto the
colour taxonomy of the cross-sensor RGB image.

\paragraph{Anomaly detection.}
Among the many algorithms for unsupervised anomaly detection on
sensor data, the Isolation Forest of \citet{liu2008isolation} has
proven robust on high-dimensional inputs.  The visualisation
introduced here is intended to complement, not replace, such
quantitative tools: it is a visual front end that surfaces
candidate regions of interest which a quantitative detector can
then score.

\section{Setup and Notation}
\label{sec:setup}

We consider a stationary array of three synchronously sampled
magnetometers, indexed throughout by $k\in\{1,2,3\}$.  The sensors
may be classical (e.g.\ fluxgate or proton-precession) or quantum
(OPM, NV centre, SQUID); the formalism that follows is agnostic to
the transduction mechanism.  No assumption is made on the relative
geometry of the three sensors beyond the requirement that they share
a common time base.  For each sensor and
each discrete time index $n$ we measure the three Cartesian
components of the local magnetic-field vector,
$b_{x,k}[n]$, $b_{y,k}[n]$, $b_{z,k}[n]$, sampled at a common rate
$\fs$.  All subsequent processing operates on the rotation-invariant
scalar magnitude
\begin{equation}
    \Bmag_{k}[n] \;=\; \sqrt{b_{x,k}^{2}[n] + b_{y,k}^{2}[n]
                              + b_{z,k}^{2}[n]}\,.
    \label{eq:Bmag}
\end{equation}
Working with $\Bmag$ rather than the individual vector components
removes any ambiguity associated with sensor-frame orientation:
$\Bmag$ is a true scalar invariant of the local field.  The price is
a loss of phase information between vector components, which the
present construction does not exploit.

We assume only that the three streams are simultaneously available
and gap-free over the analysis interval; in practice short dropouts
can be repaired by linear interpolation or by trimming all three
sensors to their common temporal support.  No high-pass or low-pass
filtering is applied prior to the STFT, so that the colour structure
of the resulting image reflects the raw spectral content of the
data.

\section{Mathematical Framework}
\label{sec:method}

The cross-sensor RGB spectrogram is constructed in five stages:
short-time Fourier analysis, computation of the power spectrum,
optional logarithmic compression, per-channel min--max normalisation,
and stacking into the channels of a single image.  We describe each
stage in turn.

\subsection{Short-time Fourier transform}
\label{sec:stft}

Let $w[n]$ be a real, length-$N$ analysis window and $H$ the hop
size between consecutive frames.  The discrete STFT of sensor $k$ at
frame index $m$ and discrete angular frequency $\omega$ is
\begin{equation}
    \STFT_{k}[m,\omega]
        \;=\;
    \sum_{n=0}^{N-1} \Bmag_{k}[n + mH]\, w[n]\, e^{-j\omega n}\,.
    \label{eq:stft}
\end{equation}
We adopt the Hann window
\begin{equation}
    w[n] \;=\; \tfrac{1}{2}\!\left(1 - \cos\!\left(\tfrac{2\pi n}{N-1}\right)\right),
    \qquad n = 0,\dots,N-1\,,
    \label{eq:hann}
\end{equation}
which provides a good compromise between main-lobe width and
side-lobe rejection \citep{harris1978windows}.  The construction
that follows is independent of this choice; any well-behaved
window with comparable properties may be substituted.

\subsection{Power spectrum}
\label{sec:power}

The associated power spectrum at frame $m$ is
\begin{equation}
    P_{k}[m,\omega] \;=\; \bigl|\STFT_{k}[m,\omega]\bigr|^{2}\,.
    \label{eq:power}
\end{equation}
The frequency resolution of the analysis is fixed by the window
length and the sampling rate,
\begin{equation}
    \Delta f \;=\; \frac{\fs}{N},
    \label{eq:df}
\end{equation}
and the time hop is
\begin{equation}
    H \;=\; \lfloor N(1-\rho)\rfloor,
    \label{eq:hop}
\end{equation}
where $\rho \in [0,1)$ is the chosen fractional overlap between
consecutive frames.  Larger $N$ improves frequency resolution at the
cost of time localisation; larger $\rho$ improves the temporal
density of frames at the cost of redundancy between adjacent
columns.  These two parameters define the time--frequency
``aspect ratio'' of the resulting image and must be chosen against
the typical duration and bandwidth of the features one wishes to
expose.

\subsection{Logarithmic compression}
\label{sec:db}

The dynamic range of $P_{k}$ is typically very large, spanning many
orders of magnitude between quiet and active bands.  To bring it
within the perceptual range of a display we apply an optional
logarithmic compression,
\begin{equation}
    P_{k}^{\mathrm{dB}}[m,\omega]
    \;=\;
    10\,\log_{10}\!\bigl(\max(P_{k}[m,\omega],\,\epsilon)\bigr),
    \label{eq:db}
\end{equation}
where the floor $\epsilon > 0$ guards against numerical underflow.
A purely linear variant simply skips
Equation~\eqref{eq:db}; the choice between the two is a matter of
which features one wishes to emphasise (the linear scaling
emphasises the brightest cells, the decibel scaling renders the
quiet cells visible at the cost of compressing the bright ones).

\subsection{Per-channel normalisation}
\label{sec:norm}

To form an RGB image each channel must be confined to $[0,1]$.  We
apply a min--max normalisation independently to each sensor:
\begin{equation}
    \tilde{P}_{k}[m,\omega]
        \;=\;
    \frac{P_{k}^{\star}[m,\omega] - \displaystyle\min_{m,\omega}P_{k}^{\star}}
         {\displaystyle\max_{m,\omega}P_{k}^{\star}
             - \displaystyle\min_{m,\omega}P_{k}^{\star}}
        \;\in\; [0,1],
    \label{eq:norm}
\end{equation}
where $P_{k}^{\star}$ stands for either $P_{k}$ or $P_{k}^{\mathrm{dB}}$
depending on the chosen scaling, and the extrema are taken over the
entire time--frequency support of the analysis interval.  The choice
between per-channel and joint normalisation is significant and
deserves comment.  A \emph{joint} normalisation, in which a single
$(\min,\max)$ pair is computed across all three sensors, would
preserve the absolute amplitude difference between sensors and
would render any sensor with anomalously low overall gain as a
dark channel.  The per-channel choice instead emphasises
\emph{shape} agreement: each sensor is rescaled into the same
$[0,1]$ range so that the final image highlights coherent
\emph{structure} in the time--frequency plane rather than coherent
amplitude.  In an anomaly-discovery setting, where the analyst does
not yet know which features matter, the per-channel choice tends to
be more informative; in a calibration-monitoring setting, where
absolute gain matters, the joint choice should be preferred.  The
two views are easily generated side by side from the same underlying
spectrograms.

\subsection{RGB fusion}
\label{sec:fusion}

The cross-sensor RGB image is the rank-three tensor
\begin{equation}
    \mathbf{I}[m,\omega]
    \;=\;
    \bigl(\tilde{P}_{1}[m,\omega],\;
          \tilde{P}_{2}[m,\omega],\;
          \tilde{P}_{3}[m,\omega]\bigr),
    \label{eq:rgb}
\end{equation}
displayed with sensor~1 on the red channel, sensor~2 on the green
channel, and sensor~3 on the blue channel.  The assignment of
sensors to colour channels is arbitrary up to a permutation; once
fixed, it must be communicated alongside the image so that the
colour code can be interpreted.  When the spectrograms of the three
sensors do not share an identical $(\text{frequency},\text{time})$
shape---for instance because one sensor lost a few samples to a
dropout---all three are first cropped to the smallest common
rectangle prior to stacking.

The resulting tensor $\mathbf{I}[m,\omega]$ is a colour image whose
horizontal axis is time, whose vertical axis is frequency, and whose
chroma at each pixel encodes the agreement of the three sensors at
the corresponding time--frequency cell.  This is the essential
output of the construction; everything that follows in
Section~\ref{sec:signatures} is a structured way of reading
$\mathbf{I}$.

\subsection{Long-window low-frequency variant}
\label{sec:lowf}

The construction above is parametrised by the analysis-window length
$N$ and the overlap $\rho$.  When the features of interest live
predominantly at low frequencies---for example, in the ULF band
relevant to geomagnetic micropulsations of class Pc3--Pc5
\citep{jacobs1964classification}---the same construction can be
re-instantiated with a substantially longer window $N_{\mathrm{lf}}
\gg N$ and a higher overlap $\rho_{\mathrm{lf}}$ approaching unity.
By Equation~\eqref{eq:df}, the longer window pushes the frequency
resolution $\Delta f = \fs/N_{\mathrm{lf}}$ down to whatever
resolution is required to separate the bands of interest, while the
higher overlap restores the temporal density of frames that the
longer window would otherwise discard.  The image is then cropped
along the frequency axis to the band of interest, $0 \le f \le
f_{\max}^{\mathrm{lf}}$, before being passed through stages
\ref{sec:db}--\ref{sec:fusion}.  We refer to the result as the
\emph{long-window} or \emph{low-frequency} variant of the
cross-sensor RGB spectrogram.  All other steps---scalar magnitude,
power, optional logarithm, per-channel normalisation, RGB
stacking---are unchanged.

\subsection{Algorithmic summary}
\label{sec:summary}

The full pipeline can be stated as a sequence of operations on each
sensor stream:

\begin{enumerate}
    \item For each $k$, form the scalar magnitude $\Bmag_{k}$ via
          Equation~\eqref{eq:Bmag}.
    \item Compute the STFT $\STFT_{k}[m,\omega]$ via
          Equation~\eqref{eq:stft} with a Hann window of length $N$
          and overlap fraction $\rho$.
    \item Form the power spectrum $P_{k}[m,\omega]$ via
          Equation~\eqref{eq:power}.
    \item Apply the optional logarithmic compression of
          Equation~\eqref{eq:db} to obtain $P_{k}^{\mathrm{dB}}$.
    \item Normalise each channel independently to $[0,1]$ via
          Equation~\eqref{eq:norm}, obtaining $\tilde{P}_{k}$.
    \item Crop all three normalised spectrograms to their smallest
          common $(\text{frequency},\text{time})$ rectangle.
    \item Stack the three cropped channels as
          $(\tilde{P}_{1},\tilde{P}_{2},\tilde{P}_{3})$ to obtain
          the cross-sensor RGB image $\mathbf{I}[m,\omega]$.
\end{enumerate}

The low-frequency variant inserts a frequency-axis crop to
$0 \le f \le f_{\max}^{\mathrm{lf}}$ between steps 5 and 6, and
substitutes a longer window $N_{\mathrm{lf}}$ and a higher overlap
$\rho_{\mathrm{lf}}$ in step 2.

\section{Anomaly Signatures in Cross-Sensor RGB}
\label{sec:signatures}

The diagnostic value of the cross-sensor image is that the three
sensors are presented as colour primaries, so that any pixel encodes
both the location of spectral energy in the time--frequency plane
\emph{and} the agreement of the three sensors at that point.  We
distinguish four broad classes of signature.

\subsection{Achromatic regions: coherent broadband energy}
\label{sec:white}

A pixel for which $\tilde{P}_{1}\!\approx\!\tilde{P}_{2}\!\approx\!\tilde{P}_{3}$
appears grey, with luminance proportional to the common value.
Bright achromatic regions correspond to time--frequency cells in
which all three sensors carry comparable normalised power.  Two
physically distinct sources can produce this pattern:
\begin{itemize}
    \item \emph{Genuine ambient signal.} A spatially extended
          magnetic disturbance---a geomagnetic storm, a Pc-class
          micropulsation, the magnetic signature of a large
          conductive body in the vicinity of the array---couples to
          all three sensors with similar magnitude and is therefore
          rendered in neutral tones.
    \item \emph{Site-wide common-mode noise.} An EMI source located
          far from any individual sensor, or a vibration mode that
          excites all three mounts equally, produces the same
          neutral signature.  Achromatic regions are therefore
          necessary but not sufficient for ``signal''; they must be
          cross-checked against ancillary monitoring channels
          before being attributed to the environment as opposed to
          the installation.
\end{itemize}

\subsection{Primary-colour patches: single-sensor anomalies}
\label{sec:primary}

A saturated red, green, or blue patch indicates that the
corresponding sensor alone carries spectral energy in that
time--frequency cell.  Common causes include local EMI from
equipment mounted close to that sensor, loose connectors or
microphonic pick-up in one cable run, temporary saturation of one
digitiser channel, and mechanical resonances of one mount.
A primary-colour patch is essentially never of ambient origin and
is the strongest visual signal that the affected sensor requires
maintenance.

\subsection{Secondary-colour patches: pairwise coherence}
\label{sec:secondary}

A pixel that is bright in two channels and dark in the third
appears in a secondary colour: yellow ($R+G$, sensor~3 absent),
magenta ($R+B$, sensor~2 absent), or cyan ($G+B$, sensor~1 absent).
Such pixels indicate that two sensors share a spectral feature that
the third does not see.  Physically this most often corresponds to
a near-field source that is closer to two of the three sensors than
to the third, so that the geometric attenuation of the source's
field at the third sensor renders that channel invisible to the
feature.  Secondary-colour patches are therefore the most direct
visual cue for \emph{spatially asymmetric} anomalies in a
three-element array, and the secondary colour identifies which two
sensors observe the source.

\subsection{Temporal colour drift}
\label{sec:drift}

A region in which the colour balance shifts \emph{slowly} along the
time axis, while the spectral structure remains the same, is
characteristic of progressive sensor decalibration or thermal drift.
Because the per-channel normalisation in Equation~\eqref{eq:norm}
acts globally over the analysis interval, a slow drift in the gain
of one sensor stretches or compresses its dynamic range relative to
the others, producing a gradual colour cast across the image.  This
is one of the few signatures that the per-channel normalisation
\emph{enhances} relative to a joint normalisation.

\subsection{Low-frequency bands}
\label{sec:lowfbands}

In the low-frequency variant the relevant frequency axis is
dominated by the geomagnetic micropulsation taxonomy of
\citet{jacobs1964classification}: Pc5 ($\sim$1.7--6.7\,\si{mHz}),
Pc4 ($\sim$6.7--22\,\si{mHz}), Pc3 ($\sim$22--100\,\si{mHz}), and
the lower edge of Pc2.  Achromatic horizontal stripes within these
bands are signatures of natural ambient activity, while saturated
stripes localised to one or two sensors indicate
installation-side contamination of the same band.  This distinction
is operationally important whenever the science target lives in the
ULF window, since a sensor-side stripe can mimic a faint ambient
band in any single-sensor view.

\subsection{Quantum-sensor-specific signatures}
\label{sec:quantumsig}

When the triad consists of quantum magnetometers, several additional
anomaly classes become relevant:
\begin{itemize}
    \item \emph{Spin-projection noise floor.}  In an OPM or NV-centre
          sensor operating near the spin-projection limit, the noise
          power spectral density is white and set by the atom number
          or NV-ensemble size.  If all three sensors are
          quantum-limited, the noise floor appears as a uniform,
          low-level achromatic background in the cross-sensor image.
          A single sensor whose noise floor is elevated (e.g.\ due
          to reduced optical pumping efficiency) will instead show a
          faint primary-colour haze across the spectrum.
    \item \emph{Quantum jumps and flux quanta.}  In a SQUID
          magnetometer, a single flux-quantum slip produces a
          discrete step in the output.  In the STFT this manifests
          as a broadband impulsive column localised to one sensor and
          rendered as a vertical primary-colour stripe.
    \item \emph{$T_{1}$ relaxation transients.}  An NV-centre sensor
          undergoing anomalous spin-lattice relaxation emits a
          broadband burst that decays on the $T_{1}$ time scale.
          Such a transient appears as a primary-colour patch whose
          vertical extent spans the full bandwidth and whose
          horizontal extent reflects $T_{1}$.
    \item \emph{Photon shot noise.}  The detection stage of an OPM
          contributes Poissonian photon noise that is spectrally
          flat.  When it dominates in one sensor but not the others,
          the corresponding channel shows a uniform primary-colour
          wash---the shot-noise analogue of the EMI stripe described
          in Section~\ref{sec:primary}.
\end{itemize}
In each case the colour code of the cross-sensor image immediately
identifies the affected sensor and distinguishes the quantum-noise
anomaly from an ambient magnetic signal, which would appear
achromatic.

\subsection{Colour--anomaly taxonomy}
\label{sec:taxonomy}

Table~\ref{tab:taxonomy} summarises the mapping from observed
colour to physical interpretation.  In practice the analyst
proceeds from the brightest non-grey regions of the cross-sensor
image down to the faintest, ranking each by colour, time--frequency
location, and duration.

\begin{table}[h]
    \centering
    \begin{tabular}{lll}
        \toprule
        Observed colour      & Sensor agreement       & Physical interpretation \\
        \midrule
        White / bright grey  & All three sensors agree & Coherent broadband signal: \\
                             &                        & ambient activity or site-wide noise \\
        Dark / black         & No power in any sensor & Quiet band \\
        Pure red             & Sensor~1 only          & Sensor-1 fault, EMI, or mount resonance \\
        Pure green           & Sensor~2 only          & Sensor-2 fault, EMI, or mount resonance \\
        Pure blue            & Sensor~3 only          & Sensor-3 fault, EMI, or mount resonance \\
        Yellow ($R{+}G$)     & Sensors 1, 2           & Asymmetric source nearer to 1 \& 2 \\
        Magenta ($R{+}B$)    & Sensors 1, 3           & Asymmetric source nearer to 1 \& 3 \\
        Cyan ($G{+}B$)       & Sensors 2, 3           & Asymmetric source nearer to 2 \& 3 \\
        Slow colour drift    & Time-varying ratio     & Sensor decalibration or thermal drift \\
        \midrule
        \multicolumn{3}{l}{\emph{Additional signatures for quantum magnetometers}} \\
        \midrule
        Achromatic floor     & All three at same      & Quantum-limited (spin-projection) \\
                             & low level              & noise floor \\
        Primary-colour haze  & One sensor elevated    & Degraded quantum efficiency in \\
                             &                        & one channel \\
        Vertical primary     & Impulsive, one sensor  & Flux-quantum jump (SQUID) or \\
         stripe              &                        & $T_{1}$ relaxation burst (NV) \\
        \bottomrule
    \end{tabular}
    \caption{Cross-sensor RGB colour--anomaly taxonomy. The same
             colour code applies to both the broadband and the
             low-frequency variants of the construction.}
    \label{tab:taxonomy}
\end{table}

\section{Discussion}
\label{sec:discussion}

\subsection{Relationship to classical coherence}

The standard quantitative measure of agreement between two sensors
at a given frequency is the magnitude-squared coherence
\begin{equation}
    \gamma_{ij}^{2}(f)
    \;=\;
    \frac{|G_{ij}(f)|^{2}}{G_{ii}(f)\,G_{jj}(f)},
    \label{eq:coh}
\end{equation}
where $G_{ij}$ denotes the cross spectral density of sensors $i$ and
$j$.  The cross-sensor RGB image does not estimate $\gamma^{2}$
explicitly, and in particular it discards phase information; what
it provides instead is a non-parametric, three-way visual surrogate
that is computed from power spectra alone and that requires no
ensemble averaging.  The two views are complementary: the
quantitative coherence is the right tool when one wishes to test a
specific frequency band against a threshold, while the cross-sensor
RGB is the right tool when one does not yet know which bands
deserve scrutiny.

A second formal connection is to the singular-value decomposition of
the $3 \times 3$ cross-spectral matrix at each $(m,\omega)$ cell:
fully coherent activity across the three sensors corresponds to a
rank-one matrix and is rendered as an achromatic pixel; rank-two
behaviour corresponds to secondary colours; rank-three (mutually
incoherent) behaviour produces saturated primary colours that vary
from pixel to pixel.  The cross-sensor RGB image can therefore be
read as a qualitative, pixelwise rank visualisation of the
cross-spectral matrix.

\subsection{Per-channel versus joint normalisation}

The per-channel normalisation of Equation~\eqref{eq:norm} preserves
\emph{shape} information at the cost of \emph{amplitude} information.
A sensor that quietly loses a fraction of its absolute gain across
the analysis interval will be rescaled back into $[0,1]$ and will
appear no different from its peers, except possibly through a slow
temporal colour drift (Section~\ref{sec:drift}).  Conversely, a
joint normalisation would render that sensor as a dark channel and
would make the gain loss obvious.  The per-channel choice is more
useful for anomaly \emph{discovery}, because it places the three
sensors on an equal visual footing and prevents one over- or
under-gained sensor from dominating the image; the joint choice is
more useful for amplitude calibration monitoring.  Both views can
be generated from the same underlying $P_{k}^{\star}$ at negligible
additional cost, and we recommend that any practical deployment of
the construction generate the two side by side.

\subsection{Window-length sensitivity}

By Equation~\eqref{eq:df}, the choice of $N$ controls the
trade-off between time and frequency resolution.  Short windows
localise events tightly in time at the cost of frequency
resolution; long windows do the opposite.  Real features in
$\mathbf{I}[m,\omega]$ tend to persist across a range of $N$, with
their visual extent trading time for frequency in the expected way;
analysis artefacts and sub-window transients tend to vanish as $N$
grows.  A practical recipe is therefore to compute the
construction at several values of $N$ spanning roughly a decade and
to retain only those features that survive the sweep.  This
multi-resolution check is the cross-sensor analogue of the
multi-window strategy widely used in single-sensor STFT analysis.

\subsection{Quantum noise floors and the standard quantum limit}

For a triad of quantum magnetometers the cross-sensor RGB image
offers a visual test of whether each sensor is operating at its
fundamental noise limit.  The standard quantum limit (SQL) for an
ensemble of $N_{\mathrm{at}}$ spin-$\tfrac{1}{2}$ atoms sets the
magnetic-field sensitivity at
$\delta B_{\mathrm{SQL}} \propto 1/\sqrt{N_{\mathrm{at}}\,T_{2}}$,
where $T_{2}$ is the spin coherence time
\citep{degen2017quantum}.  When all three sensors reach the SQL, the
per-channel normalisation renders the noise floor as a uniform
achromatic background at a common level.  A sensor that fails to
reach the SQL---because of excess technical noise, degraded optical
pumping, or a shortened $T_{2}$---will have a higher noise floor
than its peers, and after per-channel normalisation the excess power
will appear as a faint primary-colour haze.  The cross-sensor image
therefore functions as a quick, non-parametric ``SQL health check''
that complements the quantitative Allan-deviation or noise-spectral-density measurements typically used to characterise
quantum sensor performance.

\subsection{Limitations}

The construction inherits the limitations of any STFT-based
analysis: fixed time--frequency resolution per window, leakage
controlled by the choice of window, and an implicit assumption of
local stationarity within each frame.  Beyond these, the
cross-sensor RGB visualisation specifically:
\begin{itemize}
    \item is restricted to exactly three sensors, since RGB has
          three channels; arrays with four or more sensors require
          either an alternative colour space (e.g., CMYK) or a
          rotating selection of sensor triplets;
    \item discards phase information by working from
          $|\STFT_{k}|^{2}$, and therefore cannot distinguish
          in-phase from out-of-phase pairwise coherence;
    \item is intended as a visual aid and not as a quantitative
          detector; it complements but does not replace algorithms
          such as Isolation Forest \citep{liu2008isolation};
    \item assumes that the three sensors are sampled synchronously
          and at the same rate; resampling to a common grid is
          required if this assumption fails.
\end{itemize}

\section{Conclusion}
\label{sec:conclusion}

We have introduced the cross-sensor RGB spectrogram as a purely
methodological construction: a deliberately simple visualisation
that compresses the short-time Fourier power of three concurrent
magnetometers into a single colour image.  The construction has
been formalised
(Equations~\ref{eq:Bmag}--\ref{eq:rgb}), accompanied by an explicit
account of its time--frequency resolution
(Equations~\ref{eq:df}--\ref{eq:hop}), and equipped with a
colour-to-physics taxonomy (Table~\ref{tab:taxonomy}) that
distinguishes coherent broadband features, single-sensor faults,
asymmetric pairwise sources, and slow drift.  A long-window
low-frequency variant has been described for resolving features in
the ULF band.

Several extensions are natural.  First, the qualitative colour
taxonomy can be promoted to a quantitative segmentation by
classifying each pixel of $\mathbf{I}[m,\omega]$ in colour space and
flagging those whose chroma exceeds a tunable threshold.  Second,
the visualisation can be coupled to existing unsupervised anomaly
detectors so that algorithmic anomalies are overlaid on the
cross-sensor image for analyst review.  Third, a complementary view
that uses joint normalisation, or that substitutes the
magnitude-squared coherence $\gamma_{ij}^{2}$ of
Equation~\eqref{eq:coh} for the raw power spectra, would carry phase
and absolute-amplitude information into the colour channels.
Finally, the same construction can be applied to gradiometric
quantities such as $\Bmag_{i}-\Bmag_{j}$ to expose spatial gradients
in colour form.  For quantum sensor arrays, an extension that
incorporates the known quantum noise floor into the normalisation
stage---for instance by subtracting the expected SQL spectral density
before normalising---could sharpen the distinction between
quantum-limited and technically limited operation.

The cross-sensor RGB spectrogram does not aspire to replace any of
the precision tools in the multi-magnetometer pipeline.  It is a
single chart that an analyst can read in seconds and that surfaces,
in colour, the kinds of anomaly that would otherwise demand a
careful side-by-side comparison of three monochrome power spectra.
In a domain where the cost of missing a sensor fault or of confusing
an installation-induced line for an ambient signal is high, that is
already a useful place to start.

\section*{Code Availability}

A reference implementation of the construction described in this paper,
including the full seven-step pipeline, a Matplotlib renderer, and a
self-contained synthetic demonstration, is openly available at

\begin{center}
    \url{https://github.com/manasp21/cross-sensor-rgb-spectrograms}
\end{center}

The repository contains no measurement data; all examples run on
numerically synthesised streams.

\bibliographystyle{plainnat}
\bibliography{references}

@book{oppenheim1999discrete,
  author    = {Oppenheim, Alan V. and Schafer, Ronald W.},
  title     = {Discrete-Time Signal Processing},
  edition   = {2nd},
  year      = {1999},
  publisher = {Prentice Hall},
  address   = {Upper Saddle River, NJ}
}

@book{cohen1995time,
  author    = {Cohen, Leon},
  title     = {Time-Frequency Analysis},
  year      = {1995},
  publisher = {Prentice Hall},
  address   = {Englewood Cliffs, NJ}
}

@article{allen1977unified,
  author  = {Allen, Jont B. and Rabiner, Lawrence R.},
  title   = {A Unified Approach to Short-Time {F}ourier Analysis and Synthesis},
  journal = {Proceedings of the IEEE},
  volume  = {65},
  number  = {11},
  pages   = {1558--1564},
  year    = {1977}
}

@article{harris1978windows,
  author  = {Harris, Fredric J.},
  title   = {On the Use of Windows for Harmonic Analysis with the Discrete {F}ourier Transform},
  journal = {Proceedings of the IEEE},
  volume  = {66},
  number  = {1},
  pages   = {51--83},
  year    = {1978}
}

@article{welch1967psd,
  author  = {Welch, Peter D.},
  title   = {The Use of Fast {F}ourier Transform for the Estimation of Power Spectra: A Method Based on Time Averaging over Short, Modified Periodograms},
  journal = {IEEE Transactions on Audio and Electroacoustics},
  volume  = {15},
  number  = {2},
  pages   = {70--73},
  year    = {1967}
}

@book{gonzalez2018digital,
  author    = {Gonzalez, Rafael C. and Woods, Richard E.},
  title     = {Digital Image Processing},
  edition   = {4th},
  year      = {2018},
  publisher = {Pearson},
  address   = {New York}
}

@article{jacobs1964classification,
  author  = {Jacobs, J. A. and Kato, Y. and Matsushita, S. and Troitskaya, V. A.},
  title   = {Classification of Geomagnetic Micropulsations},
  journal = {Journal of Geophysical Research},
  volume  = {69},
  number  = {1},
  pages   = {180--181},
  year    = {1964}
}

@inproceedings{liu2008isolation,
  author    = {Liu, Fei Tony and Ting, Kai Ming and Zhou, Zhi-Hua},
  title     = {Isolation Forest},
  booktitle = {Proceedings of the 8th IEEE International Conference on Data Mining (ICDM)},
  pages     = {413--422},
  year      = {2008}
}

@article{budker2007optical,
  author  = {Budker, Dmitry and Romalis, Michael},
  title   = {Optical Magnetometry},
  journal = {Nature Physics},
  volume  = {3},
  number  = {4},
  pages   = {227--234},
  year    = {2007}
}

@article{degen2017quantum,
  author  = {Degen, C. L. and Reinhard, F. and Cappellaro, P.},
  title   = {Quantum Sensing},
  journal = {Reviews of Modern Physics},
  volume  = {89},
  number  = {3},
  pages   = {035002},
  year    = {2017}
}

\end{document}